\makeatletter\providecommand\declare@file@substitution[2]{}
\declare@file@substitution{revtex4-1.cls}{revtex4-2.cls}
\let\old@LaTeX\LaTeX
\documentclass[trackchanges,twocolumn,twocolappendix,tighten]{aastex631}
\ifpdf\pdfoutput=1\fi
\let\LaTeX\old@LaTeX
\makeatother
\renewcommand{\today}{2025 December 5} 
\usepackage{amsmath,xspace,tabularx}
\usepackage[all]{hypcap}
\usepackage{xparse}

\usepackage[angle-symbol-over-decimal,unit-mode=text,
table-alignment-mode=format,table-number-alignment=center,
uncertainty-mode=separate,print-unity-mantissa=false,
separate-uncertainty-units=single,
range-units=single,range-phrase={\text{--}},list-units=single]{siunitx}

\hypersetup{bookmarksnumbered=true,bookmarksopen=true,
  pdfauthor=Xuchen Lin
}

\IfFileExists{latexml.sty}
  {\usepackage{latexml}}
  {\newif\iflatexml\latexmlfalse}
\ProvideDocumentCommand\published{m}{}
\ifxetex%
  \usepackage{xeCJK}
  \usepackage{unicode-math}
    [range="002F]
    [range={cal/Latin,bfcal/Latin},StylisticSet=1]
    [range={cal/latin,bfcal/latin}]
\else
  \usepackage[T1]{fontenc}
  \iflatexml\else
    \IfFileExists{newtx.sty}{%
      \usepackage{newtx}
    }{%
      \usepackage{newtxtext,newtxmath}
    }
    \usepackage{CJK}
  \fi
  \AtBeginDocument{\DeclareSIUnit[quantity-product={}]\degree{\text{\textdegree}}}
\fi
\AtBeginDocument{\DeclareSIUnit[quantity-product={},list-units=repeat]\percent{\text{\%}}}
\ProvideDocumentEnvironment{CJK*}{m m}{}{}
\makeatletter
\AtBeginDocument{
  \newcommand*{\cjkname}[3][gbsn]{#2 (\begin{CJK*}{UTF8}{#1}#3\ignorespacesafterend\end{CJK*})}
  \@ifpackageloaded{CJK}{}{
    \@ifpackageloaded{xeCJK}{}{
      \iflatexml\else
        \renewcommand*{\cjkname}[3][]{#2}
      \fi
    }
  }
}
\usepackage{etoolbox}
\newcount\replace@count
\newcommand*{\@checkappendixparam}[1]{%
  \def\@checkappendixparamtmp{#1}%
  \ifx\Hy@param\@checkappendixparamtmp
    \let\Hy@param\Hy@appendixstring
  \fi
}
\apptocmd{\appendix}{%
  \patchcmd{\hyper@makecurrent}{%
    \edef\Hy@param{#1}%
  }{%
    \edef\Hy@param{#1}%
    \@checkappendixparam{subsection}%
    \@checkappendixparam{subsubsection}%
  }{}{}%
}{}{}
\patchcmd{\HyOrg@appendix}{\onecolumngrid}{}{}{}
\hypersetup{
  linkcolor=blue,citecolor=blue,
  filecolor=blue,urlcolor=blue
}
\pretocmd{\maketitle}{
  \hypersetup{urlcolor=black}
}{}{}
\apptocmd{\maketitle}{
  \hypersetup{urlcolor=blue}
}{}{}
\patchcmd{\NAT@citex}
{\@citea\NAT@hyper@{%
    \NAT@nmfmt{\NAT@nm}%
    \hyper@natlinkbreak{\NAT@aysep\NAT@spacechar}{\@citeb\@extra@b@citeb}%
    \NAT@date}}
{\@citea\NAT@nmfmt{\NAT@nm}%
  \NAT@aysep\NAT@spacechar\NAT@hyper@{\NAT@date}}{}{}
\patchcmd{\NAT@citex}
{\@citea\NAT@hyper@{%
    \NAT@nmfmt{\NAT@nm}%
    \hyper@natlinkbreak{\NAT@spacechar\NAT@@open\if*#1*\else#1\NAT@spacechar\fi}%
    {\@citeb\@extra@b@citeb}%
    \NAT@date}}
{\@citea\NAT@nmfmt{\NAT@nm}%
  \NAT@spacechar\NAT@@open\if*#1*\else#1\NAT@spacechar\fi\NAT@hyper@{\NAT@date}}
{}{}
\iflatexml\providecommand{\onecolumngrid}{}\fi
\AtBeginEnvironment{thebibliography}{%
  \phantomsection%
  \addcontentsline{toc}{section}{\refname}%
}
\patchcmd{\thebibliography}{%
  \onecolumngrid}{}{}{}
\patchcmd{\thebibliography}{%
  \twocolumngrid}{}{}{}
\patchcmd{\thebibliography}{%
  \textwidth}{\linewidth}{}{}
\patchcmd{\thebibliography}{%
  \raggedright\small}{%
  \footnotesize%
}{}{}
\patchcmd{\thebibliography}{%
  \baselineskip=13pt plus 1pt}{%
  \baselineskip=9pt plus 1pt%
}{}{}
\patchcmd{\thebibliography}{%
  \baselineskip=13pt plus 1pt}{%
  \baselineskip=9pt plus 1pt%
}{}{}
  \patchcmd{\@makecaption}{\small}{\footnotesize}{}{}
  \advance\replace@count-1
\ifnum\replace@count>0\repeat
\patchcmd{\@facility}{%
  \large}{}{}{}
\patchcmd{\@facilities}{%
  \large}{}{}{}
\patchcmd{\@software}{%
  \large}{}{}{}
\patchcmd{\sec@upcase}{%
  \uppercase}{%
  \MakeTextUppercase%
}{}{}
\renewcommand\ltx@foottext[2]{%
  \begingroup
  \expandafter\ltx@make@current@footnote\expandafter{\@mpfn}{#1}%
  \@footnotetext{#2}
  \endgroup
}%
\renewcommand\@makefntext[1]{%
  \parindent 1em%
  \baselineskip 8.5\p@%
  \noindent
  \ifnum\c@footnote<100%
    \hb@xt@1.25em{\@makefnmark\hss}%
  \else%
    \hbox{\@makefnmark\ }%
  \fi#1%
}
\patchcmd{\acknowledgments}{%
  \begin{internallinenumbers}}{%
    \ifnumlines\begin{internallinenumbers}\fi%
      }{}{}
      \patchcmd{\endacknowledgments}{%
    \end{internallinenumbers}}{%
    \ifnumlines\end{internallinenumbers}\fi%
}{}{}
\patchcmd{\enddeluxetable}{%
  \null
}{}{}{}
\renewcommand\ApjSectionpenalty{\@M}
\renewcommand\aapr{\ref@jnl{A\&ARv}}
\renewcommand\nat{\ref@jnl{Natur}}
\makeatother

\makeatletter
\renewcommand\ion[2]{\texorpdfstring{\text{#1\,\textsc{\@roman{#2}}}}{#1 \@roman{#2}}}
\makeatother
\newcommand*{\Hi}{\ion{H}{1}\xspace}
\newcommand*{\sbHi}[1][]{\tsb{\Hi\if\relax\detokenize{#1}\relax\else,#1\fi}\xspace}
\newcommand*{\tsb}[1]{\ensuremath{_{\text{#1}}}}
\newcommand*{\Htwo}{H\tsb{2}\xspace}

\newcommand*{\SfG}{S\textsuperscript{4}G\xspace}
\newcommand*{\fov}{FoV\xspace}
\newcommand*{\fwhm}{\text{FWHM}\xspace}
\newcommand*{\deltaS}{\ensuremath{I_{18}}\xspace}
\newcommand*{\logSratOuter}{\ensuremath{\log(S_{18}/S_{19})}\xspace}
\newcommand*{\logSratInner}{\ensuremath{\log(S_{19}/S_{20})}\xspace}
\newcommand*{\rs}{\ensuremath{r\tsb{s}}\xspace}
\newcommand*{\Ri}{\ensuremath{R\sbHi}\xspace}
\newcommand*{\Rooi}{\ensuremath{R\sbHi[001]}\xspace}
\newcommand*{\zIB}{\ensuremath{z_{18}}\xspace}
\newcommand*{\MHi}{\ensuremath{M\sbHi}\xspace}
\newcommand*{\MHiDiemer}{\ensuremath{M\sbHi[D18,L08]}\xspace}
\newcommand*{\MHifaceon}{\ensuremath{M\sbHi[face-on]}\xspace}
\newcommand*{\Mste}{\ensuremath{M_\ast}\xspace}
\newcommand*{\Mhalo}{\ensuremath{M\tsb{halo}}\xspace}
\newcommand*{\MBH}{\ensuremath{M\tsb{BH}}\xspace}
\newcommand*{\MBHdot}{\ensuremath{\dot{M}\tsb{BH}}\xspace}
\newcommand*{\NHi}{\ensuremath{N\sbHi}\xspace}
\newcommand*{\fHi}{\ensuremath{f\sbHi}\xspace}

\newcommand*{\sfr}{\ensuremath{\text{SFR}}\xspace}
\newcommand*{\ssfr}{\ensuremath{\text{sSFR}}\xspace}

\newcommand*{\simufield}[1]{\texttt{#1}}
\newcommand*{\softwarename}[1]{\texttt{#1}}

\makeatletter
\@ifpackageloaded{unicode-math}{%
  \newcommand{\inc}{\increment}
}{%
  \iflatexml
    \newcommand{\inc}{\Delta}
  \else
    \IfPackageAtLeastTF{newtx}{2023/11/14}{%
      \newcommand{\inc}{\laplace}
    }{%
      \newcommand{\inc}{\Delta}
    }
  \fi
}\makeatother

\newlength{\myhalfimgsize}
\newcommand*\myplotone[1]{%
  \centering
  \leavevmode
  \includegraphics[width={%
        \ifdim\textwidth=\linewidth%
          2\myhalfimgsize%
        \else\ifdim\linewidth>\myhalfimgsize%
            \myhalfimgsize%
          \else%
            \linewidth
          \fi\fi}]{#1}%
}%

\ProvideDocumentCommand{\comment}{mO{cyan}+m}{%
  {\color{#2}\textbf{#1}: #3}%
}

\DeclareSIUnit\smallh{\ensuremath{h}}
\DeclareSIUnit\yr{yr}
\DeclareSIUnit\Gyr{\giga\yr}
\DeclareSIUnit\Myr{\mega\yr}
\DeclareSIUnit\G{G}
\DeclareSIUnit\uG{\micro\G}
\DeclareSIUnit\pc{pc}
\DeclareSIUnit\Msun{M\ensuremath{_\odot}}
\DeclareSIUnit\kpc{\kilo\pc}
\DeclareSIUnit\Mpc{\mega\pc}
\DeclareSIUnit\dex{dex}
\DeclareSIUnit\deg{deg}
\DeclareSIUnit\erg{erg}
\DeclareSIUnit\mag{mag}
\DeclareSIUnit\Jy{Jy}
\DeclareSIUnit\mJy{\milli\Jy}
\DeclareSIUnit\beam{beam}
\DeclareSIUnit\pixel{pixel}

\newcommand*{\autoeqref}[1]{\hyperref[#1]{Equation~(\ref*{#1})}}
\iflatexml
  
  \providecommand{\textLambda}{\ensuremath{\Lambda}}
\else
  \usepackage{textalpha}
\fi

\ProvideDocumentCommand\unit{om}{\si[#1]{#2}}
\ProvideDocumentCommand\qty{omm}{\SI[#1]{#2}{#3}}
\ProvideDocumentCommand\qtylist{omm}{\SIlist[#1]{#2}{#3}}
\ProvideDocumentCommand\qtyrange{ommm}{\SIrange[#1]{#2}{#3}{#4}}
\iflatexml\else
\fi

\graphicspath{{fig/}{anc/}}

\defcitealias{2023ApJ...944..102W}{W23}
\defcitealias{2008AJ....136.2782L}{L08}
\defcitealias{2018ApJS..238...33D}{D18}

\shorttitle{FEASTS versus Simulations}
\shortauthors{Lin et al.}

\received{2025 September 21}
\revised{2025 December 4}
\accepted{2025 December 5}
\published{\today}

\submitjournal{The Astrophysical Journal}

\begin{document}

\title{
  FEASTS Compared with Simulations:
  Abnormally Irregular and Extended \Hi Morphologies at a Column Density of \qty[number-mode=text,reset-text-series=false]{1e18}{\per\cm\squared} in TNG50 and Auriga
}

\correspondingauthor{\cjkname{Jing Wang}{王菁}}
\email{jwang\_astro@pku.edu.cn}


\author[0000-0002-4250-2709]{\cjkname{Xuchen Lin}{林旭辰}}
\affiliation{Department of Astronomy, School of Physics, Peking University, Beijing 100871, People's Republic of China}

\author[0000-0002-6593-8820]{\cjkname{Jing Wang}{王菁}}
\affiliation{Kavli Institute for Astronomy and Astrophysics, Peking University, Beijing 100871, People's Republic of China}

\author{Guinevere Kauffmann}
\affiliation{Max-Planck Institut f\"ur Astrophysik, Karl-Schwarzschild-Stra\ss{}e 1, D-85741 Garching, Germany}

\author[0000-0001-5976-4599]{Volker Springel}
\affiliation{Max-Planck Institut f\"ur Astrophysik, Karl-Schwarzschild-Stra\ss{}e 1, D-85741 Garching, Germany}

\author[0000-0003-3308-2420]{R\"udiger Pakmor}
\affiliation{Max-Planck Institut f\"ur Astrophysik, Karl-Schwarzschild-Stra\ss{}e 1, D-85741 Garching, Germany}

\begin{abstract}
  With new atomic-hydrogen (\Hi) observations of FAST Extended Atlas of Selected Targets Survey (FEASTS), we present the first statistical comparison of \Hi morphology between observations and cosmological simulations, focusing on low--column density (\qty{\sim1e18}{\per\cm\squared}) regions of Milky Way--like central galaxies.
  We select a 330-galaxy sample from IllustrisTNG50 (TNG50) matched to 33 FEASTS galaxies by stellar and \Hi masses, and mock observe them to the FAST resolution and depth at corresponding inclinations and distances for a fair comparison.
  In contrast to FEASTS, abnormally irregular and extended morphology is found in more than one-third of TNG50 galaxies, especially those massive and \Hi poor.
  Stellar feedback is the property that most significantly correlates with the \Hi morphological deviation from observations, although these deviations mostly occur at a high stellar or black-hole mass.
  These results indicate that in TNG50, stellar feedback significantly influences the \Hi morphology at \qty{\sim1e18}{\per\cm\squared}, while active galactic nucleus (AGN) feedback has not so direct a role as expected.
  With an additional sample from Auriga, we find that the magnetic field may help \Hi to be more regular in its morphology, while improving the mass resolution does not alleviate the discrepancy from observation.
  This study reveals the potential of constraining future simulations of galaxies by observing low--column density \Hi.
\end{abstract}

\section{Introduction}
\label{sec:intro}
In the Lambda cold dark matter (\textLambda{}CDM) cosmological context, the baryon cycle is a key component regulating galaxy evolution \citep{2020ARA&A..58..363P}.
Around galaxies, the circumgalactic medium (CGM) provides the reservoir of gas accretion and serves as the passage through which gas recycles back to the interstellar medium (ISM) after being kicked out by feedback processes \citep{2017ARA&A..55..389T}.
The multiphase CGM has temperatures and densities spanning orders of magnitude \citep{2023ARA&A..61..131F}, setting a complex environment for the evolution of ISM and galaxies.

This interface between CGM and ISM offers important constraints for galaxy-formation models \citep{2023ARA&A..61..473C}, but a comprehensive and quantified description of it requires multiwavelength observations.
The low--column density, cool gas in CGM and outer ISM has been primarily observed with absorption lines \citep{2017ASSL..430..167C,2017ASSL..430..117L}.
The aggregation of single--line of sight absorption measurements provides the column-density distribution function \citep[e.g.,][]{2005MNRAS.363..479P,2021ApJ...923...50F} and, after associating them with nearby possible absorbers, averaged spatial dependence \citep[e.g.,][]{2023MNRAS.523..676W,2024MNRAS.527.3494W}.
Due to the sparsity of bright quasars \citep{2024arXiv240900174F}, the interpretation has caveats of systematic source-to-source variation and degeneracy in geometry and kinematics.

Another possibility is directly mapping the \qty{21}{\cm} emission from neutral atomic hydrogen (\Hi).
With recent advances in radio astronomy equipments and technologies, a relatively low \Hi column density (\NHi) of \qty{\sim5e17}{\per\cm\squared} has been reached for an intermediate-sized sample at a relatively low time cost by programs like FAST Extended Atlas of Selected Targets Survey \citep[FEASTS;][]{2023ApJ...944..102W,2025ApJ...980...25W} and MeerKAT \Hi Observations of Nearby Galactic Objects: Observing Southern Emitters \citep[MHONGOOSE;][]{2024A&A...688A.109D}.
\Hi in unperturbed galaxies is found to continuously extend to this \NHi as a relatively smooth, dynamically cold rotating disk \citep{2024A&A...688A.109D,2024ApJ...968...48W}, which can reach half the virial radius \citep{2025ApJ...980...25W}.

\Hi gas properties have not been taken as a~priori constraints in major cosmological simulations \citep{2023ARA&A..61..473C}.
The comparison between simulations' predictions and observations is usually conducted on a few physically motivated simple properties, such as \NHi distributions \citep{2012MNRAS.421.2809V,2017MNRAS.464.4204C}, \Hi mass function \citep{2020MNRAS.497..146D,2025A&A...699A..14W}, and \Hi--stellar mass relation \citep{2017MNRAS.464.4204C,2019MNRAS.487.1529D}, of which no simulation suite so far has fully reproduced all.

Meanwhile, the morphological comparison of \Hi disks has been focused on high-\NHi regions.
Compared with previous \Hi interferometric surveys with an \NHi sensitivity of \qty[input-comparators=\gtrsim]{\gtrsim1e19}{\per\cm\squared}, \Hi disks in Evolution and Assembly of GaLaxies and their Environments (EAGLE) \citep{2016MNRAS.456.1115B} and The Next Generation Illustris (IllustrisTNG) \citep{2023MNRAS.521.5645G,2023ApJ...957L..19S} are deficient in \Hi in the central region, likely caused by overly strong stellar or active galactic nucleus (AGN) feedbacks.
Beyond the central region, the tight observational relation between the characteristic \Hi radius \Ri at a surface density of \qty{1}{\Msun\per\pc\squared} and \Hi mass \citep{2016MNRAS.460.2143W} is roughly reproduced in simulations but with a systematic offset depending on the recipe partitioning molecular and atomic hydrogen \citep{2017MNRAS.466.3859M,2019MNRAS.487.1529D}.

As an ecosystem, the star-forming activity in the inner region of galaxies is regulated by the active gas inflow and outflow at a lower \NHi (\qty[input-comparators=\gtrsim]{\gtrsim5e17}{\per\cm\squared} in this paper), where many simulations hint that \Hi structures show filamentary and clumpy features \citep[e.g.,][]{2023MNRAS.518.5754R}, with details possibly dependent on resolution \citep{2019MNRAS.482L..85V}, magnetic field \citep{2020MNRAS.498.2391N}, and radiation field \citep{2019MNRAS.490.1518O}.
Directly quantify and compare the morphology of the simulated low-\NHi features with observations may help to constrain related physical processes, which has not been possible until recently.
With seven almost perfectly edge-on galaxies, \citet{2025ApJ...984...15Y} found \Hi disks much thinner vertically than simulated by Illustris or IllustrisTNG \citep{2016MNRAS.462.3751K,2019MNRAS.486.4686K}, measured at $\NHi=\qty{1e18}{\per\cm\squared}$.
Compared with five Milky Way (MW)--like galaxies observed with MeerKAT, 25 simulated IllustrisTNG and Fire-2 galaxies in a similar stellar-mass range have a higher fraction of $\NHi<\qty{1e20}{\per\cm\squared}$ spaxels, possibly contributing to their irregular morphology \citep{2025A&A...697A..86M}.

In this paper, we conduct a statistical comparison of \Hi morphology between FEASTS and simulations at a low \NHi.
Leveraging the fact that the FEASTS sample is biased toward \Hi-rich massive galaxies, we select a sample of 33 MW-like galaxies from FEASTS, and compile a simulation sample controlling for both stellar and \Hi mass.
In simulations, \Hi gas has generally been estimated in a postprocessing fashion \citep{2023ARA&A..61..473C}, because the adopted subgrid physics recipe cannot trace low-temperature gas on the fly due to limited resolutions and computational costs \citetext{\citealp[e.g.,][]{2015MNRAS.446..521S,2017MNRAS.464.4204C,2017MNRAS.466.3859M,2018ApJS..238...33D,2019MNRAS.483.5334S}; but also see \citealp{2020MNRAS.497..146D}}.
The choice of models for partitioning \Hi and molecular hydrogen (\Htwo) has been found to have non-negligible influence on their mass and distribution \citep{2018ApJS..238...33D,2023MNRAS.521.5645G,2025A&A...697A..86M}.
By focussing on low-\NHi region where the \Htwo-fraction is negligible, we alleviate the uncertainty introduced by the partition model.

This paper is structured as follows.
We present the details on sample selection and data processing in \autoref{sec:sample}, and introduce three morphological parameters in \autoref{sec:morph_para}.
In \autoref{sec:results}, we analyze the distribution of morphological parameters, which is different between FEASTS and simulations;
for simulations, we identify the major galactic parameters that correlate with the \Hi morphological difference from the real Universe.
These correlations are likely artificial and not physical, but potentially point to the most relevant processes to blame for, as discussed in \autoref{sec:discussion}.
We summarize in \autoref{sec:summary}.
For the observed data, we assume a \textLambda{}CDM with $\Omega\tsb{m}=0.3$, $\Omega\tsb{\textLambda}=0.7$, and $h=0.7$, and adopt the stellar initial mass function of \citet{2003ApJ...598.1076K}.

\begin{figure}
  \centering
  \includegraphics{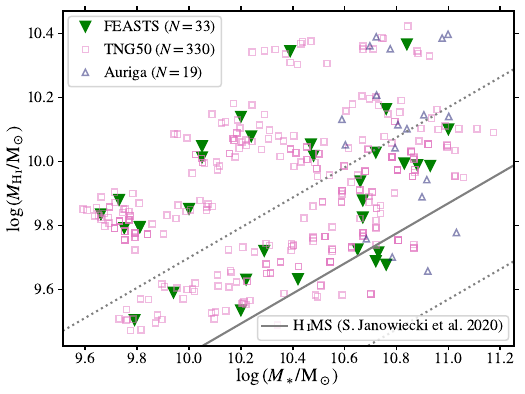}
  \caption{
    The \Hi mass (\MHi) and stellar mass (\Mste) of our observational sample (filled green triangles), TNG50 matched sample (empty pink squares), and the Auriga sample (empty blue triangles).
    The \MHi--\Mste relation of star-forming galaxies in xGASS is plotted as the gray line, with the \qty{\sim0.3}{\dex} scatter indicated by dotted lines.
  }
  \label{fig:sample}
\end{figure}


\begin{figure*}
  \footnotesize
  {
    \centering
    \includegraphics{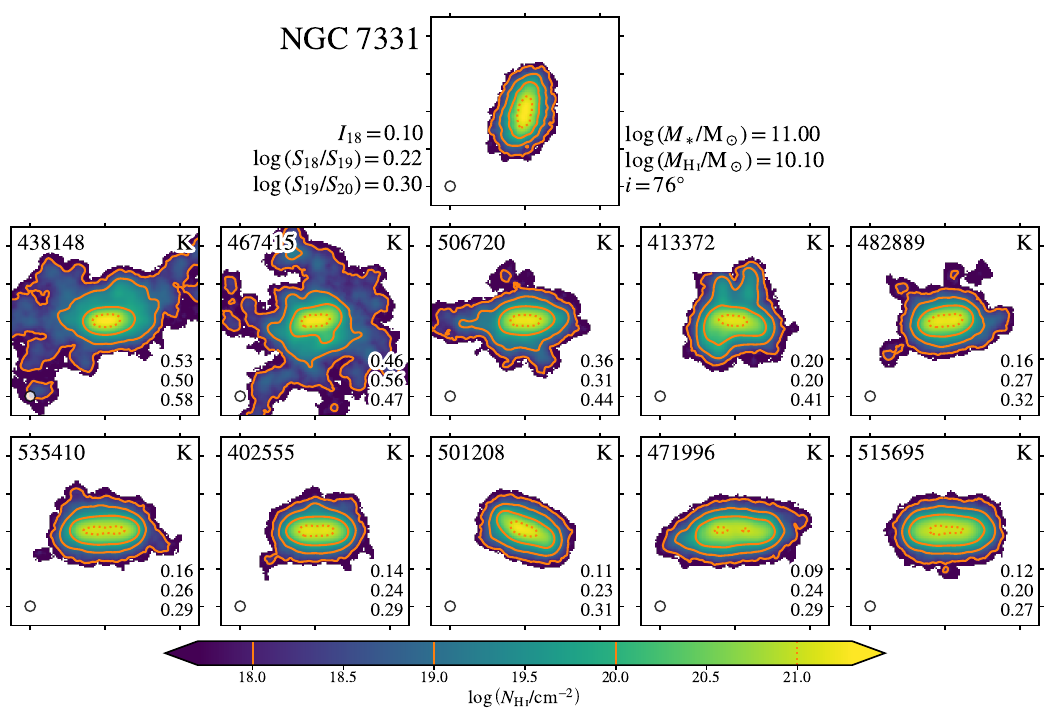}
    \par
  }
  \caption{
    The \Hi moment-0 atlas of FEASTS galaxies (the single panel above) and their distance-matched TNG50 cubes sorted by the values of morphological parameters in descending order (ten panels below).
    In the full Figure Set, FEASTS galaxies are sorted by \Mste in descending order.
    All panels have a sidelength of \qty{250}{\kpc}, and the beam size of \ang{;3.24;} is indicated by the circle at the lower left corner.
    On the left of FEASTS moment-0 map, we list the NGC name and morphological parameters;
    on the right, we list the stellar and \Hi masses and the inclination.
    In the TNG panels, we list their subhalo ID, AGN-feedback mode (``K'' for kinetic mode and ``T'' for thermal mode), and morphological parameters.
    Morphological parameters listed in the atlas are not corrected for the distance.
    Contours are plotted at $\NHi=\text{\qtylist{1e18;1e19;1e20}{\per\cm\squared}}$ (solid lines) and \qty{1e21}{\per\cm\squared} (dotted lines).
  }
  (The complete figure set (33 images) is available as the ancillary file to this arXiv submission.)
  \label{fig:atlas}
\end{figure*}

\section{Data and Sample}
\label{sec:sample}
\subsection{Observational Sample from FEASTS}
The \emph{observational sample} consists of 33 FEASTS galaxies (green triangles in \autoref{fig:sample}), taken from a parent sample of 55 galaxies that have been observed by FEASTS from September 2021 to June 2024 \citep[also see][]{2025ApJ...980...25W}.
An atlas is given in the Figure Set associated with \autoref{fig:atlas}, with NGC~7331 shown at the top of \autoref{fig:atlas} as an example.
We first set a stellar-mass threshold of $\Mste>\qty[parse-numbers=false]{10^{9.5}}{\Msun}$ and then exclude eight satellite galaxies.
A sample galaxy is classified as a satellite if it is not the $K_s$-band brightest galaxy of its group according to the group catalog of \citet{2017ApJ...843...16K}.
We further remove one galaxy that is significantly \Hi poor (NGC~4826, with \Hi mass fraction $\fHi=\MHi/\Mste=10^{-1.75}$).
All of these 33 galaxies are located on or above the typical \Hi--stellar mass relation of star-forming galaxies in the Extended \emph{GALEX} Arecibo SDSS Survey (xGASS) \citep[gray line]{2020MNRAS.493.1982J}, biased toward \Hi-rich galaxies for the convenience of observation.

Values of \Mste, star-formation rate (\sfr), and distance are taken from the $z=0$ Multiwavelength Galaxy Synthesis \citep[$z$0MGS;][]{2019ApJS..244...24L} catalog.
Their \Mste and \sfr estimations are calibrated against the \emph{GALEX}--SDSS--\emph{WISE} Legacy Catalog \citep[GSWLC;][]{2016ApJS..227....2S,2018ApJ...859...11S}.
For the 33 galaxies in our sample, they calculated \sfr from the integrated fluxes in the \emph{Galaxy Evolution Explorer} (\emph{GALEX}) far-ultraviolet (FUV) band and the \emph{Wide-field Infrared Survey Explorer} (\emph{WISE}) W4 band, and estimated \Mste from the \emph{WISE} W1-band flux with the mass-to-light ratio predicted from the ratio between \sfr and W1 flux.
Distances are compiled from multiple sources, with a typical uncertainty of \qty{0.1}{\dex}.
The \Hi mass \MHi is calculated from the integrated \Hi \qty{21}{\cm} emission-line flux of FEASTS cube within the source-detected region under an optically-thin assumption.
We calculate the inclination from the \qty{3.6}{\um} isophotal axis ratio by the Spitzer Survey of Stellar Structure in Galaxies \citep[\SfG;][]{2010PASP..122.1397S}.%
\footnote{
  Four galaxies are not included in \SfG.
  We take their inclinations from \citet[NGC~6384]{2003A&A...412...45P}, \citet[NGC~891]{2007AJ....134.1019O}, and \citet[NGC~925 and~7331]{2008AJ....136.2648D}.
}

FEASTS observations were carried out using basket-weaving on-the-fly scans, and were reduced with a dedicated pipeline \citep{2023ApJ...944..102W}.
The field of view (\fov) is typically chosen to be $\ang{1}\times\ang{1}$, successfully enclosing \Hi structures of the target galaxy in the majority of cases.
The product cubes have a beam of \ang{;3.24} \fwhm (ranging between \qtylist{6.5;24.3}{\kpc}), a pixel size of \ang{;;30}, and a channel width of \qty{1.610}{\km\per\s}.
The median noise rms level is \qty{1.0}{\mJy\per\beam}, corresponding to a $3\sigma$ \Hi column density detection limit of \qty[parse-numbers=false]{10^{17.7}}{\per\cm\squared} assuming a line width of \qty{20}{\km\per\s}.
Due to the gridding process, the noise has a Gaussian spatial autocorrelation with $\fwhm=\ang{;1.45}$.

The source-finding process of FEASTS uses \softwarename{SoFiA} \citep{2015MNRAS.448.1922S,2021MNRAS.506.3962W} with the parameters listed in \citet{2024ApJ...968...48W}.
Some galaxies at an early stage of merger have not experienced significant morphological disturbance, but are connect with its counterpart in the \Hi cube due to the limited spatial resolution.
We use the 3D-deblending method of \citet{2025ApJ...980..157H} to separate the \Hi fluxes of these early-merging pairs, as done by \citet{2025ApJ...980...25W}.

\subsection{TNG50 Simulation and Parent Sample}
IllustrisTNG \citep[TNG;][]{2018MNRAS.480.5113M,2018MNRAS.477.1206N,2018MNRAS.475..624N,2018MNRAS.475..648P,2018MNRAS.475..676S} is a set of cosmological simulations built upon the magnetohydrodynamical moving-mesh simulation code \softwarename{Arepo} \citep{2010MNRAS.401..791S,2019ascl.soft09010S,2020ApJS..248...32W}.
As the follow-up of Illustris \citep{2014MNRAS.445..175G,2014MNRAS.444.1518V,2014Natur.509..177V,2015MNRAS.452..575S}, TNG (1)~includes magnetic fields, (2)~adopts a new ``kinetic'' AGN feedback model at low accretion rates that randomly injects momentum to surroundings, (3)~applies isotropic stellar winds with thermal energy and a velocity floor, and (4)~updates the stellar evolution and gas chemical enrichment processes \citep{2017MNRAS.465.3291W,2018MNRAS.473.4077P}.
In this paper, we focus on comparing FEASTS \Hi observations with TNG50-1 \citep{2019MNRAS.490.3234N,2019MNRAS.490.3196P}, the highest-resolution volume of TNG, which has a box sidelength of \qty{\sim35}{\Mpc\per\smallh} and a baryonic mass resolution of \qty{8.5e4}{\Msun}.

With similar selection criteria as for the observational sample, we compile an initial sample consisting of all 875 TNG50 central galaxies with stellar mass \Mste between \iflatexml$10^{9.5}$ and \qty[parse-numbers=false]{10^{11.2}}{\Msun}\else\qtylist[parse-numbers=false]{10^{9.5};10^{11.2}}{\Msun}\fi{} and a catalog $\MHi>\qty[parse-numbers=false]{10^{8.7}}{\Msun}$.
We use \Mste values from \citet{2021MNRAS.500.3957E}%
\footnote{
  \url{https://www.tng-project.org/data/docs/specifications/\#sec5p}
}
with \qty{30}{\kpc} apertures, as in \citet{2019MNRAS.490.3234N} and \citet{2019MNRAS.490.3196P}.
During sample selection, we take the catalog \MHi (hereafter \MHiDiemer ) from \citet[hereafter \citetalias{2018ApJS..238...33D}]{2018ApJS..238...33D},%
\footnote{
  \url{https://www.tng-project.org/data/docs/specifications/\#sec5i}
}
calculated with the empirical \Hi--\Htwo transition model of \citet[hereafter \citetalias{2008AJ....136.2782L}]{2008AJ....136.2782L}.
The \citetalias{2008AJ....136.2782L} model is an update to the midplane pressure--based model of \citet{2006ApJ...650..933B}, which has been tested to perform better than recipes considering the ultraviolet background in \citet{2025A&A...697A..86M}.
Later, after conducting mock \Hi observations, we will update the values of \MHi.

Moreover, the specific \sfr ($\ssfr=\sfr/\Mste$) values are required to be larger than \qty[parse-numbers=false]{10^{-11.5}}{\per\yr}, by which we avoid the quenched galaxies in TNG50 with overpredicted \Hi masses \citep{2022ApJ...941..205M}.
We use \sfr values averaged across the last \qty{100}{\Myr} with \qty{30}{\kpc} apertures \citep{2019MNRAS.490.3196P}.%
\footnote{
  \url{https://www.tng-project.org/data/docs/specifications/\#sec5b}
}
This selection step reduces the initial sample to 828 galaxies, which make the \emph{TNG50 parent sample}.

\subsection{\Hi Postprocessing}
\label{ssec:postprocessing}

We conduct \Hi postprocessing and mock observation on TNG50 $z=0$ galaxies with a modified version%
\footnote{
  Based on \softwarename{MARTINI} v2.1.3.
  The modified version is publicly available at \url{https://github.com/CastleStar14654/martini/releases/tag/Lin25}.
}
of \softwarename{MARTINI} \citep{2019MNRAS.482..821O,2024JOSS....9.6860O}, a Python package generating \Hi emission-line cubes from simulation snapshots.
We refine the memory usage of \softwarename{MARTINI} and revise the postprocessing code for TNG as described below.

As \citetalias{2018ApJS..238...33D}, we take the neutral-gas fractions of non-star-forming cells directly from TNG snapshots (\simufield{Neutral\-Hydrogen\-Abundance}) and calculate those of star-forming cells following \citet[see also \citealp{2019MNRAS.483.5334S}]{2003MNRAS.339..289S}, who modeled the gas as one hot phase and one cold phase in pressure equilibrium.
For \Hi--\Htwo transition, we use the empirical model of \citetalias{2008AJ....136.2782L} on a cell-by-cell basis, taking the partial pressure of cold-phase gas as the midplane pressure following \citet{2017MNRAS.466.3859M}.
We verify that if we sum up the \Hi mass within the friends-of-friends (FoF) subhalos as done by \citetalias{2018ApJS..238...33D} for the TNG50 parent sample, the results are larger than \MHiDiemer by \qty{0.038(28)}{\dex} ($3\sigma$-clipped median and scatter), but generally consistent.

\subsection{\Hi Mock Observation}
\label{ssec:parent_mock_obs}
\softwarename{MARTINI} generates the mock \Hi cube from all gas cells within the given \fov and heliocentric velocity range, instead of including only the cells within the FoF subhalo.
Considering the line broadening due to turbulence and/or physical heat, the \Hi spectrum of a single gas cell is modeled as a Gaussian function with a velocity dispersion determined from the cell internal energy, assuming the hydrodynamical equilibrium between each unresolved phases.

For the TNG50 parent sample, we set their \Hi disks as face on, and obtain their mock \Hi cube with a \fov sidelength of \qty{500}{\kpc} (\ang{2.86}) at a distance of \qty{10}{\Mpc}, spanning a velocity range of \qty{700}{\km\per\s}.
The orientation of an \Hi disk is determined from the angular momentum of \qty{30}{\percent} most-central gas cells belonging to the parent FoF group, following the default behavior of \softwarename{MARTINI}\@.
While these selected gas cells are not equivalent to the \Hi disk itself, we verify by visual inspection that for the \Hi-rich, central TNG galaxies that we select, this method successfully identifies the \Hi disk orientation in most of the cases.

The pixel size and channel width are the same as FEASTS products.
The cubes are convolved with the average of the 19-beam FAST receiver beams \citep[hereafter ``FAST beam''; see Appendix~B.2 of][]{2024ApJ...968...48W}.
We add to the cube a random Gaussian noise with the same rms level and spatial autocorrelation as those of FEASTS data, and conduct the same source finding and a similar 3D-deblending procedure.
The initial guesses for deblending are taken as the locations and velocities of satellites with $\Mste>\qty{1e7}{\Msun}$.
A satellite will be removed from initial guesses if no clear \Hi structure is associated with it, or if it is so close to the target galaxy that a successful deblending is impossible.
The \Hi-mass values (\MHifaceon) calculated from these face-on cubes after source finding and 3D deblending are similar to the halo-based ones in \autoref{ssec:postprocessing}, larger by \qty{0.001(7)}{\dex}.

\begin{figure*}
  \centering
  \includegraphics{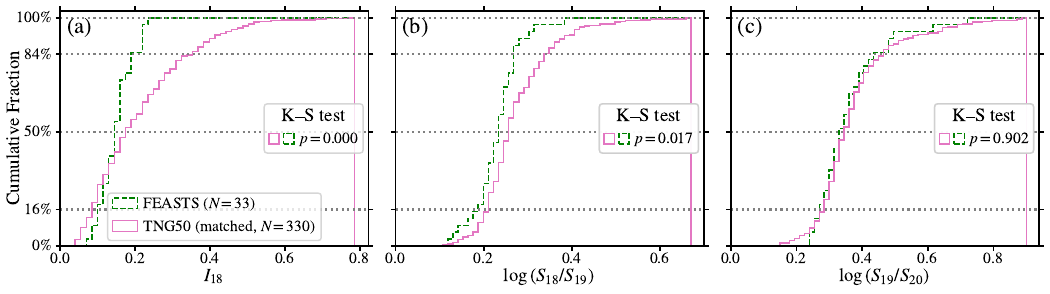}
  \caption{
    The cumulative distribution of three morphological parameters (\deltaS, \logSratOuter, and \logSratInner from left to right) and their dependence on distances $D$.
    The FEASTS observational sample and TNG50 distance-matched sample are represented by green dashed and pink solid lines, respectively.
    The $p$-value of a K--S test between the two samples is given in the legend.
  }
  \label{fig:tng_feasts_hist}
\end{figure*}

\subsection{TNG50 Matched Sample and Relative Mock \Hi Data}
We match each galaxy in the observational sample with ten TNG50 galaxies that have the most similar \Mste and \MHifaceon.
We call these 330 galaxies the \emph{TNG50 matched sample} (pink squares in \autoref{fig:sample}), serving as the control sample.
An atlas of their moment-0 maps is given as the Figure Set associated with \autoref{fig:atlas}, with the ten TNG50 galaxies matched to NGC~7331 shown in \autoref{fig:atlas} as examples.
Among these ten TNG50 galaxies, some are highly irregular (e.g., 438148 and 467415), while some have a morphology as regular as NGC~7331 (e.g., galaxies in the bottom row).
We will try to identify the condition for irregularity to occur in TNG50.

In our matching process, one TNG galaxy can be assigned to different FEASTS galaxies and be counted more than once;
the cases where the inclination difference between two FEASTS galaxies is less than \ang{10} account for 50 out of 330 galaxies.
The typical maximum difference in \MHi during the matching is \qty{0.05}{\dex}, while the difference in \Mste is three times more tolerable than in \MHi, accounting for its larger uncertainty in the observed value.
The $p$-values of Kolmogorov--Smirnov (K--S) tests are \numlist{0.73;0.90} for \Mste and \MHi, respectively, verifying the success of parameter matching.
Incidentally, the \sfr{}s (and \ssfr{s}) of these two samples also have a similar distribution with K--S test $p$-values of \num{0.76} (and \num{0.44}), although we do not take (s)\sfr as a matching parameter.

For the matched sample, we generate mock \Hi cubes in the same way as for parent sample in \autoref{ssec:parent_mock_obs}, but set the disk inclination to that of the corresponding FEASTS galaxy.
Two versions are generated, one at the corresponding FEASTS distance (\emph{distance-matched cubes}) and one at \qty{10}{\Mpc} (\emph{\qty[mode=text,reset-text-shape=false]{10}{\Mpc} cubes}).
The \fov sidelength is the maximum one of \qty{500}{\kpc} and \ang{1.5} to fully enclose the \Hi structures directly connected to central galaxies at the sensitivity depth.
Other procedures including source finding and 3D deblending are kept the same.

Additionally, according to whether the kinetic-mode AGN feedback is switched on \citep{2017MNRAS.465.3291W},%
\footnote{
  The criterion for the kinetic mode switching on is the Eddington ratio $\MBHdot/\dot{M}\tsb{Edd}$ being smaller than $\min\{0.002(\MBH/\qty{1e8}{\Msun})^2,0.1\}$, where \MBHdot is the BH accretion rate and $\dot{M}\tsb{Edd}$ is the Eddington accretion rate.
}
we separate the TNG50 matched sample into the \emph{kinetic-mode} and \emph{thermal-mode} samples.
They are labeled with a letter ``K'' or ``T'' in \autoref{fig:atlas} and the associated Figure Set.

\subsection{Auriga Simulation and Sample}
\label{ssec:auriga_sample}

Auriga \citep{2017MNRAS.467..179G,2024MNRAS.532.1814G} is a suite of cosmological ``zoom-in'' magnetohydrodynamical simulation of 40 MW-mass halos and 26 dwarf-galaxy haloes.
It is performed with \softwarename{Arepo} as TNG, but has a slightly different physics model.
Important differences include (1)~radio-mode AGN feedback modeled as randomly located ``thermal bubble,'' rather than randomly oriented kinetic jets and (2)~not setting a velocity floor for galactic winds.
The fiducial Auriga simulation (level~4) has a similar mass resolution as TNG50, and the higher--mass resolution (level~3, better by \qty{\sim1}{\dex}) and/or hydrodynamical-only version of several halos are available.

We thus take Auriga as a supplementary dataset for assessing the influences of magnetic fields and simulation mass resolution (see \autoref{sec:discussion}).
Four (or six) of Auriga level-4 galaxies have a hydrodynamical-only (high--mass resolution) counterpart for our comparison.

We also compare Auriga with TNG, checking possible effects of their differences in physics models, e.g., AGN and stellar feedback.
We conduct \Hi postprocessing and face-on mock observation as in Sections \ref{ssec:postprocessing} and~\ref{ssec:parent_mock_obs}.
We select the 19 Auriga fiducial-set (level-4) galaxies with $\Mste>\qty[parse-numbers=false]{10^{10.5}}{\Msun}$ and $\qty[parse-numbers=false]{10^{9.65}}{\Msun}<\MHi<\qty[parse-numbers=false]{10^{10.45}}{\Msun}$ as the \emph{Auriga sample} (blue triangles in \autoref{fig:sample}), which have a rather uniform coverage of the \MHi--\Mste space.

\section{Morphological Parameters}
\label{sec:morph_para}
We use three parameters to quantify the morphology of \Hi moment~0 maps, i.e., the integrated \Hi flux maps, at low column densities (\NHi).
These parameters are designed to reflect the azimuthal regularity and the relative spatial extension of low-\NHi structures, involving the contours at $\NHi=\text{\qtylist{1e18;1e19;1e20}{\per\cm\squared}}$.
If several non-connected regions are above the \qty{1e20}{\per\cm\squared} \NHi threshold, we only consider the one closest to the galaxy center.
By doing so, the \qty{1e20}{\per\cm\squared} contour is set to be a relatively simple reference point for other two contours.

The first parameter is irregularity (\deltaS), the relative area deviation of the \qty{1e18}{\per\cm\squared} \NHi contour from an ellipse with the same celestial area.
The ellipse's position angle and ellipticity are computed from the second-order moments of the region within the more-regular \qty{1e19}{\per\cm\squared} contour.
The computation is the same as that for elliptical geometric parameters in the photometry procedure of \softwarename{SExtractor} \citep{1996A&AS..117..393B} but with a uniform weighting.
A similar parameter has been used by \citet{2013MNRAS.433..270W} at a higher \NHi (\qty{7e19}{\per\cm\squared}).
The second and third parameters are similar, defined as the ratio between areas of \qtylist{1e18;1e19;1e20}{\per\cm\squared} contours.
We refer to \logSratInner and \logSratOuter as the inner and outer area ratios, respectively.
For reference, in the atlas in \autoref{fig:atlas} and the associated Figure Set, the contours of $\NHi=\text{\qtylist{1e18;1e19;1e20}{\per\cm\squared}}$ are given along with the values of \deltaS, \logSratInner, and \logSratOuter.

Our FEASTS observational sample have a wide range of distances (\qtyrange[range-phrase={ to }]{7.0}{25.9}{\Mpc}) and thus spatial resolutions, which may influence the observational morphological parameters.
This problem is alleviated by our matching the distance when conducting mock observations.
We verify that our results relating to comparing FEASTS and simulations are not influenced by distance (also see \autoref{app:sec:morph_dist_depend}).
We thus stick with the morphological parameters measured at original FEASTS distance when comparing FEASTS with simulations, and only use the \qty{10}{\Mpc} cubes when conducting analysis within simulation datasets.

\begin{figure}
  \centering
  \includegraphics{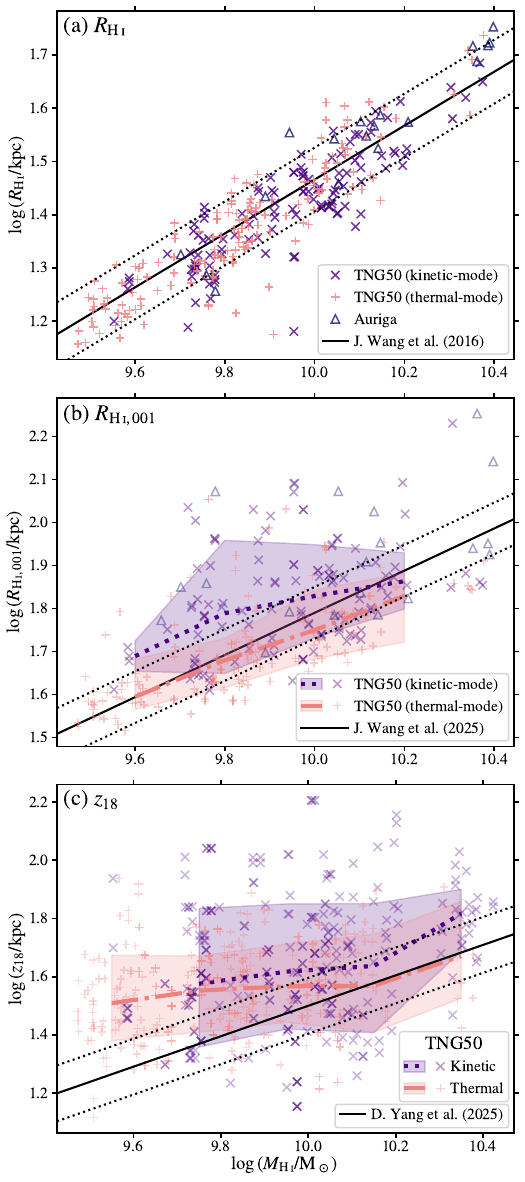}
  \caption{
    The absolute radial and vertical extensions of TNG50 \Hi disks compared with observational relations.
    (a, b)~The \Hi size--mass relation with \Ri and \Rooi, the radii at \Hi surface densities of \qtylist{1;0.01}{\Msun\per\pc\squared} after correcting for projection effects.
    Inclinations are required to be less than \ang{90}.
    Kinetic- and thermal-mode TNG50 galaxies are labeled as purple crosses and orange pluses, respectively, and fiducial Auriga galaxies are plotted as blue triangles.
    The observed relations from \citet{2016MNRAS.460.2143W,2025ApJ...980...25W} are plotted as solid lines for reference, and a scatter of \qty{\sim0.06}{\dex} is indicated by dotted lines.
    (c)~The extension \zIB perpendicular to the \Hi disk at a column density of \qty{1e18}{\per\cm\squared}, measured using edge-on mock observation.
    The observed relation from \citet{2025ApJ...984...15Y} is plotted as the solid line for reference with a scatter of \qty{\sim0.1}{\dex}.
    The binned median values and the 16th and 84th percentiles are plotted as purple dotted line and shading for kinetic mode and as orange dashed--dotted line and shading for thermal mode in panels (b) and~(c).
  }
  \label{fig:size_mass}
\end{figure}

We also make supplemental measurements of the absolute radial and vertical extension with respect to the disk plane, the characteristic \Hi disk radii (\Ri and \Rooi) at \qtylist{1;0.01}{\Msun\per\pc\squared} (\qtylist{\sim1.3e20;1.3e18}{\per\cm\squared}) and \zIB, the maximum vertical extension at \qty{1e18}{\per\cm\squared}.
Although \Ri and\Rooi provide only azimuthally averaged information and \zIB is difficult to obtain observationally due to the sparsity of edge-on galaxies, they can be easily compared with existing and future observational or simulation results.
More details of the measurement of \Ri, \Rooi, and \zIB can be found in \autoref{app:sec:size_mass}.

In this study, we focus on the low-\NHi region that has been newly detected systematically by FEASTS, therefore using three morphological parameters highlighting the shape and extension of \qty{1e18}{\per\cm\squared} \NHi contours without a strong influence from the inner, \Hi-bright part.
We do not use canonical morphological parameters that have been used for \Hi analyses \citep[e.g.,][]{2011MNRAS.416.2401H}, as they are weighted by surface brightness and thus more suitable for high-\NHi interstellar regions.
In contrast, the low-\NHi region we focus on is \qty{\sim100} times fainter and accounts for less than \qty{10}{\percent} of the total \Hi flux \citep{2005MNRAS.364.1467Z}.
Additionally, the distribution of canonical \Hi morphological parameters is significantly influenced by the existence of \Hi central holes \citep{2023MNRAS.521.5645G}, which is abnormally prevalent in TNG simulations and potentially artificially caused by the quasar-mode AGN ionization in TNG \citep{2023ApJ...957L..19S}, but beyond the scope of this paper.
We also do not discuss the presence or absence of supernova-driven \Hi holes throughout the \Hi disks, because these holes are closely associated with the high--column density star-forming regions \citep{2023ApJ...944L..22B}, and are unresolved under the resolution of the FEASTS data.


\section{Results}
\label{sec:results}
\subsection{Distribution of Morphological Parameters}
\label{ssec:para_dist}
The cumulative distribution of the three morphological parameters are shown in \autoref{fig:tng_feasts_hist} for the FEASTS observational sample (green dashed) and TNG50 matched sample (distance-matched, pink solid).
Compared to FEASTS, TNG50 galaxies have systematically higher values of \deltaS and \logSratOuter, and the distributions have a long tail toward the high end.
The low $p$-values of K--S tests confirm the significance of the difference.
More than one-third of TNG50 galaxies have a higher \deltaS than any FEASTS galaxies do, or a higher \logSratOuter than \qty{84}{\percent} of FEASTS galaxies do.
On the other hand, two samples show no significant differences in the distribution of \logSratInner.

In \autoref{fig:size_mass}, we compare TNG50 \Ri, \Rooi, and \zIB measurements with observational relations from \citet{2016MNRAS.460.2143W,2025ApJ...980...25W} and \citet{2025ApJ...984...15Y}.
The \Ri--\MHi relation is well recovered \citep[see also][]{2019MNRAS.487.1529D}, while both \Rooi and \zIB are systematically larger in TNG50 than in observations.
In this figure, we plot TNG kinetic- and thermal-mode samples separately as purple crosses and orange plusses, the distinction between which will be checked later in \autoref{ssec:para_depend}.

These results suggest that TNG50 meets difficulty in reproducing the observed low-\NHi structures and predicts more-irregular and extended \Hi structures at $\NHi\sim\qty{1e18}{\per\cm\squared}$, while the high-\NHi inner \Hi-disk structures are rather well predicted by it.
Since in TNG50 mocks, the low-\NHi regions are highly irregular, radii measured with azimuthally averaged surface density profiles are not a proper characterization for it.
In the following regarding TNG50, we mainly focus on the two morphological parameters associated with the $\NHi=\qty{1e18}{\per\cm\squared}$ contours, \deltaS and \logSratOuter.

\subsection{Morphological Correlation with Physical Properties}
\label{ssec:para_depend}
\begin{figure*}
  \centering
  \includegraphics{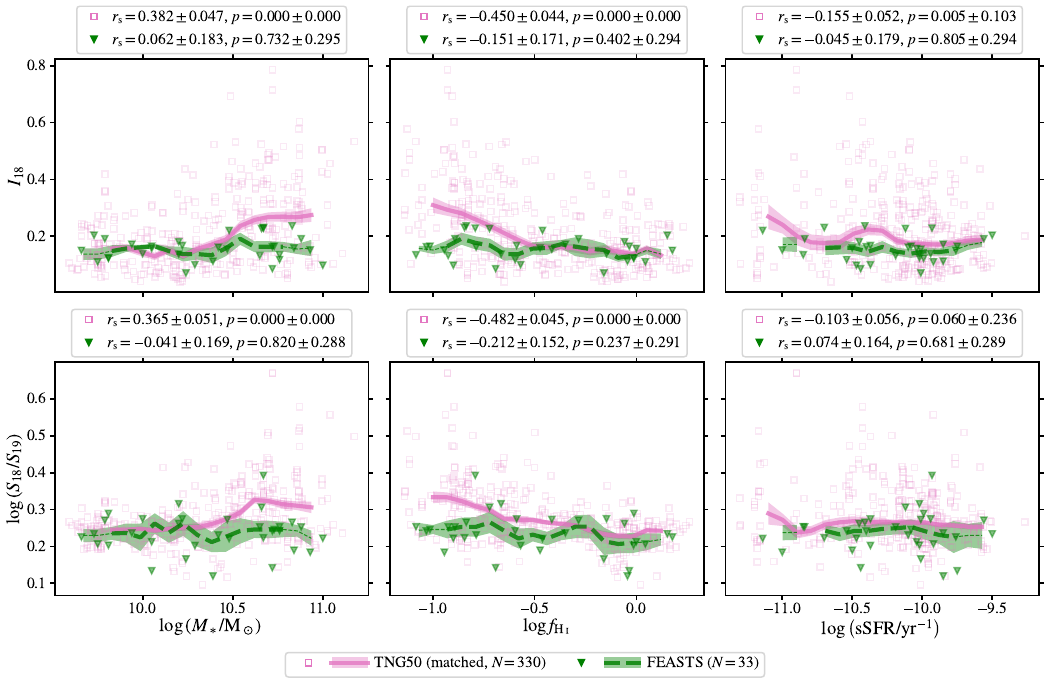}
  \caption{
    The relations between two morphological parameters at FEASTS distance (\deltaS and \logSratOuter; two rows) and three physical properties (\Mste, \fHi, and \ssfr; three columns).
    FEASTS observational sample and TNG50 matched sample are represented by green triangles and pink squares, respectively.
    The Spearman rank correlation coefficients (\rs) are listed above every panel for each sample, along with the $p$-values;
    their uncertainties are obtained with bootstrapping.
    The running median values of the morphological parameters are plotted as pink solid (TNG50) or green dashed (FEASTS) profiles.
    The widths of median filters are chosen as \qty{20}{\percent} of dynamic ranges in physical properties, i.e., \qtylist{0.32;0.28;0.41}{\dex} for \Mste, \fHi, and \ssfr, respectively, and the profile steps are \qty{5}{\percent} of the ranges.
    Only bins with at least four data points are plotted.
    We remove the incomplete bins for TNG50 profiles, and indicate them as thin lines for FEASTS\@.
    The uncertainty of median values is indicated by translucent shadings.
  }
  \label{fig:para_depend}
\end{figure*}

In this section, we present the relations between the two morphological parameters (\deltaS and \logSratOuter) and galaxy physical properties, plotted in \autoref{fig:para_depend}.
We choose three physical properties:
stellar mass $\Mste$, \Hi mass fraction $\fHi=\MHi/\Mste$, and \ssfr.
TNG50 and FEASTS galaxies are marked as pink squares and green triangles, respectively.
The Spearman rank correlation coefficient \rs for each sample is given above each panel, along with the $p$-value.
The uncertainties of \rs and $p$-value are obtained with bootstrapping.
The running median profiles are overlaid as pink solid lines (TNG50) or green dashed lines (FEASTS) to guide the eye.

The FEASTS observational sample shows no rank correlation between any of the two morphological parameters and the three physical properties.
Meanwhile, morphological parameters of the TNG50 matched sample have significant correlation with \Mste and anticorrelation with \fHi, indicating that galaxies with more-irregular or more-extended \Hi structures tend to have a large \Mste or a low \fHi in TNG50.
The high-\Mste (${\gtrsim}\qty[parse-numbers=false]{10^{10.5}}{\Msun}$) and low-\fHi (${\lesssim}10^{-0.5}$) ends are also where FEASTS and TNG50 running-median profiles diverge significantly, corresponding to a higher fraction of \Hi abnormality.
On the another hand, at the high-\Mste and low-\fHi ends, there are still a fraction of TNG50 galaxies having similar morphological parameters to those of FEASTS, statistically echoing what we find with the moment-0 maps in \autoref{fig:atlas}, i.e., both irregular and regular TNG50 galaxies can be matched to the same FEASTS galaxy.
With \ssfr, TNG50 galaxies have no morphological correlation, the same as FEASTS\@.

In the left column of \autoref{fig:para_depend}, the TNG50 running-median profiles of the two morphological parameters increase steeply at $\Mste=\qty[parse-numbers=false]{10^{10.5}}{\Msun}$, where in TNG, the kinetic-mode AGN feedback starts to operate \citep{2019MNRAS.490.3234N}, and the largest difference from observation occurs at larger \Mste's.
Correspondingly, in \autoref{fig:size_mass}, the largest deviations from observational \Rooi or \zIB relations reached by kinetic-mode galaxies are systematically higher than thermal-mode ones (see also their median profiles in panels (c) and~(d)).

\begin{figure}
  \centering
  \includegraphics{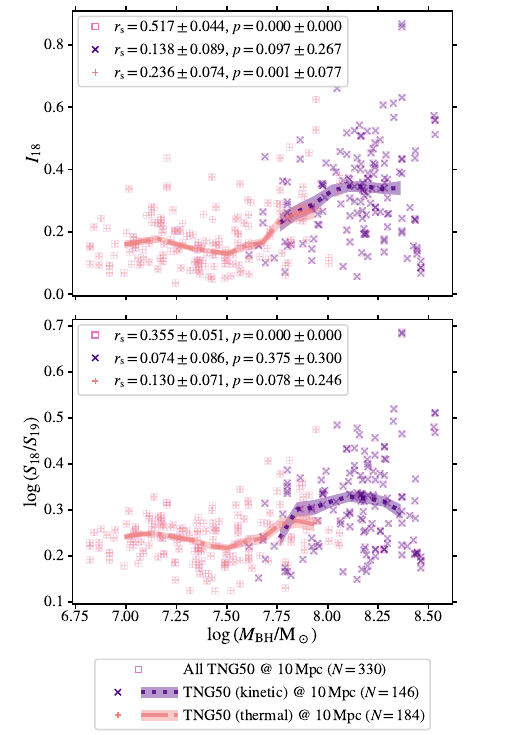}
  \caption{
    The same as \autoref{fig:para_depend}, but the physical property is the mass of BH (\MBH) and morphological parameters are measured for TNG50 galaxies at \qty{10}{\Mpc}.
    We separately plot the TNG50 kinetic-mode (purple crosses and dotted profile) and thermal-mode (orange pluses and dashed--dotted profile) samples.
  }
  \label{fig:para_depend_MBH}
\end{figure}

To check the possible influence on \Hi from the AGN feedback, we plot TNG \Hi morphological parameters against the black-hole (BH) mass \MBH in \autoref{fig:para_depend_MBH}.
Purple crosses and dotted lines represent the kinetic-mode sample, and orange pluses and dashed--dotted lines represent the thermal-mode one.
The global relations are very similar to those against \Mste, a result of the tight \MBH--\Mste relation in TNG \citep{2017MNRAS.465.3291W}.
Although kinetic-mode galaxies have significantly larger \deltaS and \logSratOuter, within the transitional \MBH range, the two subsamples do not show any discernable differences.
Possibly, the onset of kinetic-mode AGN feedback has a gentle influence on \Hi morphology, given that \MBHdot and \ssfr change smoothly after the transition \citep{2025MNRAS.543.1878L,2025MNRAS.537.3543F}.
Both thermal- and kinetic-mode AGN feedbacks may influence the bulk of \Hi gas in an indirect way by gradually modifying the CGM properties.

\subsection{The Primary Physical Property in TNG50 that Influences \Hi Morphology}
We show in \autoref{ssec:para_depend} that the \Hi morphology of TNG50 galaxies strongly correlates with \Mste, \fHi, and additionally \MBH, while that of FEASTS galaxies does not.
Such unrealistic correlations point to the artificial coupling of galactic properties in TNG50 due to simplified baryonic recipes, but it is not straightforward to identify the culprit because both \fHi and \MBH depend on \Mste.
Possible candidates include galaxy assembly history, gas accretion, in-disk gas flow, instantaneous and cumulative AGN feedback, star formation, and stellar feedback.
In this section, we use partial Spearman rank correlation to deduce the primary cause of the unrealistic \Hi morphology in TNG50.

In \autoref{fig:spearman_r}, we visualize the partial correlations between our morphological parameters (two rows of panels) and seven physical properties (\Mste, \fHi, \MHi, \MBH, BH accretion rate \MBHdot, \ssfr, and \sfr).
The three columns give the results of the full TNG50 matched sample and the thermal- and kinetic-mode samples, respectively.
Within each panel, we line variables of correlation along the vertical axis, and covariates that we control for along the horizontal axis.
The pixels are color-coded by the strength of the correlation, and those with $p<0.05$ are denoted by a ``+'' symbol to highlight their significance.

We define the \emph{primary} property as the one with all its $p$-values smaller than \num{0.05}, given as bold ``X'' symbols.
If multiple properties satisfy this requirement, we compare their partial correlations controlling for each other, and select the one with the strongest correlation.
If a property has all of its partial $|\rs|$ values within \num{0.05} of the corresponding one of the primary property, we also take it as a primary property, e.g., \MBH for \deltaS of the full \qty{10}{\Mpc} sample.
The \emph{secondary} property, indicated by bold ``+'' symbols, is the strongest one with $p<0.05$ after controlling for the primary property.

For the irregularity, \deltaS, the primary property of the whole TNG50 matched sample is \Mste, and the secondary one is total \sfr.
For the two subsamples, the primary property is total \sfr, and the secondary property is \MBH and \MHi for kinetic-mode and thermal-mode subsamples, respectively.

For the outer area ratio, \logSratOuter, the primary and secondary physical properties are always among \Hi mass (fraction) and (specific) \sfr.
The partial correlations with \sfr are more significant for the kinetic-mode sample, while \Hi masses seem to be more important for thermal-mode galaxies.
When we consider the whole sample, the quantities normalized by \Mste (\fHi and \ssfr) become the primary and secondary properties, possibly due to the wider dynamic range of \Mste than in the subsamples.

In summary, when $\Mste\gtrsim\qty[parse-numbers=false]{10^{10.5}}{\Msun}$ (or $\MBH\gtrsim\qty{1e8}{\Msun}$), the star formation plays an important role in setting all these morphological parameters, particularly dominating when the kinetic-mode AGN feedback is on.
Meanwhile, when including the remainder of the sample that does not show significant difference in \Hi morphology from observations, the \MBH (and \Mste) exhibit a stronger correlation with these morphological parameters than \sfr, implying some condition-setting or diagnosing role of the \MBH for star-formation regulations to work.
Within the full TNG50 matched sample, \MBH (and maybe also \Mste) seems to most strongly determine the contour shape at \qty{1e18}{\per\cm\squared}.
The \Hi richness most effectively regulates area ratios.

\subsection{Influences of Feedback Prescriptions}
\begin{figure*}
  \centering
  \includegraphics{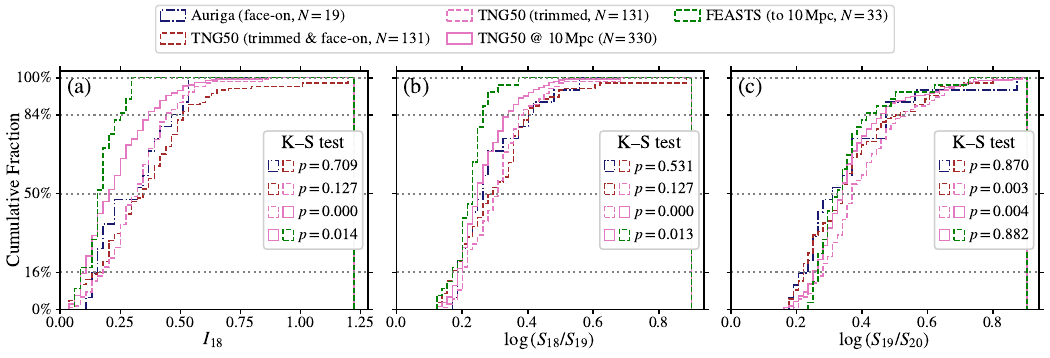}
  \caption{
    Similar to \autoref{fig:tng_feasts_hist}, but for face-on measurements of the Auriga sample (face-on, blue dashed--dotted line) and the TNG50 \qty{10}{\Mpc} sample that is trimmed to the same \Mste and \MHi range (brown dashed).
    For reference, the trimmed-but-inclination-matched TNG50 sample (pink dashed), the full TNG50 sample (pink solid), and the FEASTS \qty{10}{\Mpc} measurements (green dashed) are also plotted.
  }
  \label{fig:auriga_hist}
\end{figure*}

In \autoref{fig:auriga_hist}, we compare the distributions of \Hi morphological parameters in TNG50 and in Auriga, tentatively testing effects of their differences in physics models.
To make a fair comparison, we trim the TNG50 \qty{10}{\Mpc} sample with the same \Mste and \MHi ranges of the Auriga sample (listed in \autoref{ssec:auriga_sample}) into \num{131} galaxies, and measure the morphological parameters with face-on cubes.

The Auriga sample (blue dashed--dotted line) and the trimmed face-on TNG50 sample (brown dashed) are not significantly different in their morphological-parameter distributions.
This \Mste range of ${>}\qty[parse-numbers=false]{10^{10.5}}{\Msun}$ is where the kinetic-mode AGN feedback becomes efficient in TNG galaxies \citep{2019MNRAS.490.3234N} and where \Hi morphologies deviate the most from observations (\autoref{fig:para_depend}).
It seems that the different AGN- and/or stellar-feedback implementations in Auriga \citep{2017MNRAS.467..179G,2024MNRAS.532.1814G} also cannot produce \Hi structures as smooth and regular as observed by FEASTS\@.

\begin{figure*}
  \centering
  \includegraphics{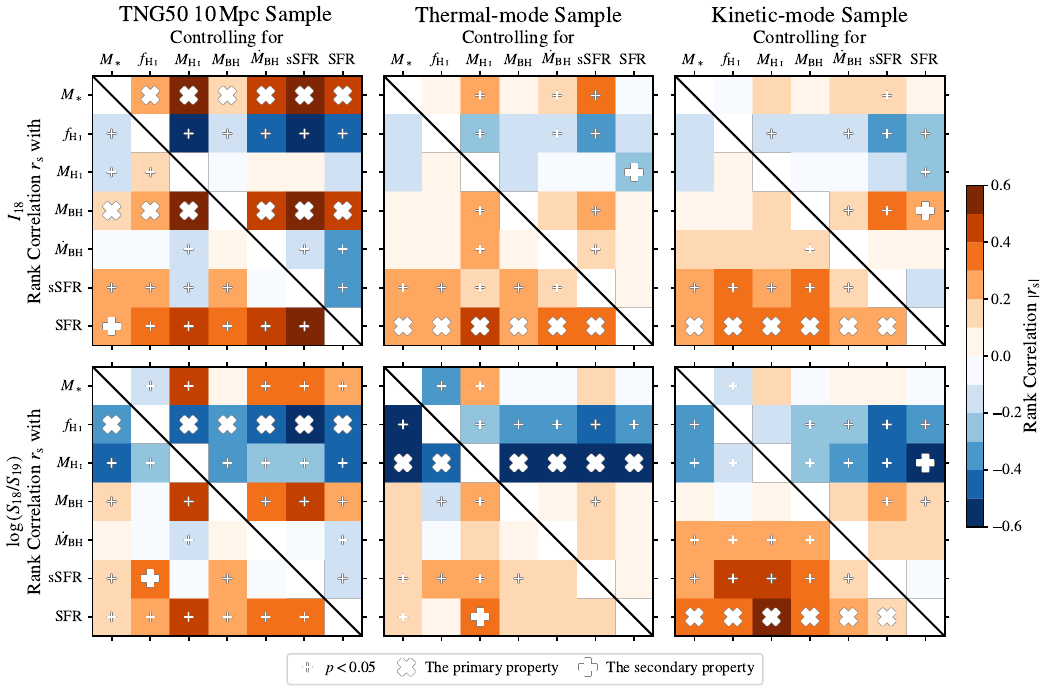}
  \caption{
    Partial Spearman rank correlation \rs between three morphological parameters (\deltaS, \logSratOuter, and \logSratInner; three rows) and seven galaxy properties (\Mste, \fHi, \MHi, \MBH, \MBHdot, \ssfr, and \sfr), indicated by colors.
    A ``+'' symbol denotes $p<0.05$, and the primary (or secondary) galaxy property is represented by bold ``X'' (or ``+'') symbols.
    Three columns represent the full TNG50 \qty{10}{\Mpc} sample and the thermal- and kinetic-mode samples, respectively.
    Within each panel, variables and covariates of the correlation are listed vertically and horizontally, respectively.
  }
  \label{fig:spearman_r}
\end{figure*}

Comparing face-on (brown dashed) and inclination-matched (pink dashed) \Hi morphological measurements of TNG50, we also find that the inclination does not significantly influence the values of \deltaS and \logSratOuter.
The face-on sample seems to have a systematically smaller value of \logSratInner than the inclination-matched sample, which might be specifically related to the vertical structure of \Hi but is beyond the main scope of this paper.
The most drastic difference is found between the TNG50 sample before (pink solid) and after (pink dashed) trimming by \Mste and \MHi, echoing the results in Figures \ref{fig:para_depend} and~\ref{fig:spearman_r} that morphological parameters strongly depend on them.

\subsection{Magnetic Field and Mass Resolution}
\begin{figure}
  \centering
  \includegraphics{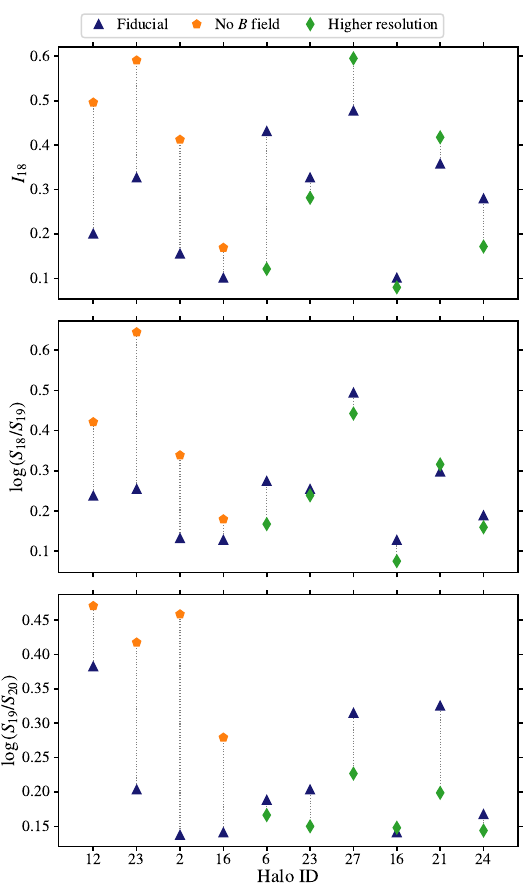}
  \caption{
    The morphological parameters of the same Auriga galaxy in different simulation sets:
    fiducial (empty blue triangles), hydrodynamical-only (i.e., without the magnetic field; orange pentagons), or high--mass resolution (green diamonds).
    The symbols of the same galaxy are linked by vertical gray dotted lines.
  }
  \label{fig:auriga_hydro_res}
\end{figure}

In \autoref{fig:auriga_hydro_res}, we compare the morphological parameters measured with different versions of Auriga.
Vertical gray dotted lines link the corresponding galaxies in the fiducial set (empty blue triangles) and in the hydrodynamical-only set (i.e., without the magnetic field; orange pentagons) or high--mass resolution set (green diamonds).

Despite the limited sample size (four), all of the hydro\-dynamical-only galaxies have a higher value of the morphological parameters than fiducial galaxies, indicating that the magnetic field helps \Hi structures be smoother and less extended, thus being closer to observations.
This effect may arise due to the supporting role of magnetic pressure in the hot CGM \citep{2020MNRAS.498.2391N} and the ISM, so that cooling gas is less apt to fragmentation.
Another possibility is that due to the confinement by magnetic field, the metal-rich bipolar outflow of gas from feedback is not well mixed with the CGM \citep{2020MNRAS.498.3125P,2021MNRAS.501.4888V}, also making the process of gas cooling into the disk less efficient.
Although the magnetic field is found not to significantly influence global galaxy properties \citep{2017MNRAS.471..144S,2020MNRAS.492.3465H}, it may still influence the local distribution of \Hi in low density regions.

Meanwhile, the high--mass resolution set and fiducial set do not show a clear trend in their differences, which fluctuate around zero with a large amplitude.
Apart from the small sample size (six), it is possible that changing the baryon mass resolution within the range of \qtylist{5e4;6e3}{\Msun} does not cause distinct and systematic differences in their \Hi structures after the images being smoothed to a \qty{\sim10}{\kpc} resolution by the FAST beam.

The literature has pointed out that spatial and mass resolutions could have a different role on \Hi morphologies \citep{2023ARA&A..61..131F}.
\citet{2019MNRAS.482L..85V} improved the spatial resolution of Auriga fiducial set to a maximum cell size of \qty{1}{\kpc}, finding that the \NHi profile becomes significantly higher and flatter at $\NHi<\qty{1e19}{\per\cm\squared}$, while the total \MHi almost does not change.
Given that the \logSratOuter of TNG50 or Auriga is already too high compared to observations, the \logSratOuter value of the ``\qty{1}{\kpc}'' simulation would be even higher, implying that simply improving the spatial resolution might not directly solve the problem of the simulated \Hi morphology.

\section{Indications for Physical Recipes in TNG50}
\label{sec:discussion}
A galaxy is an ecosystem regulated by the flow and circulation of baryonic matter \citep{2020ARA&A..58..363P}.
Cool-phase gas condenses from the CGM, getting accreted and forming the ISM disk, where it flows in to feed star formation and the central supermassive BH\@.
Their stellar and AGN feedbacks may inject baryonic matter and thermal/kinetic energy into both the ISM disk and CGM, and thus alter their thermodynamic properties, influencing ensuing rounds of gas accretion and star formation.

In this study, we find that the low--column density \Hi of the central galaxies predicted by TNG is much more filamentary, irregular, and extended than in observations.
In search for the cause for such unrealistic \Hi morphology, we analyze its correlation with galactic parameters within TNG50 in \autoref{ssec:para_depend}.
Even though the physical recipes in simulations are simplified, the simulated galactic properties are often regulated and contributed by multiple processes interfering with each other, as in the real Universe.
Notably, the simulated morphological artifacts correlate most strongly and directly with \sfr and/or total \Hi mass (\autoref{fig:spearman_r}), indicating a close link with gas accretion, star formation, and stellar feedback in TNG\@.

Nonetheless, apart from the stellar and AGN feedback models, the abnormal \Hi morphology in TNG50 may also come from the lack of physical processes that cosmological simulations have not implemented yet \citep{2023ARA&A..61..473C}, such as thermal conduction, cosmic rays, and radiation transfer, which could in particular be relevant close to the disk.

\subsection{The Dominant Role of Stellar Feedback}
The stellar feedback in TNG is implemented as isotropic galactic winds with a minimum velocity floor that are decoupled from the local ISM \citep{2018MNRAS.473.4077P}, imparting both thermal and kinetic energy to the CGM\@.
This feedback mechanism has been found responsible for redistributing baryons within the halo virial radius \citep{2024MNRAS.532.3417W}, even after the kinetic-mode AGN feedback is switched on \citep{2023MNRAS.524.5391A}, consistent with the dominating role of (s)\sfr (within the past \qty{100}{\Myr}) found in \autoref{fig:spearman_r}.

On the global scale of a galaxy, the stellar feedback could influence the \Hi morphology by producing recycling galactic fountains or by changing the CGM environments.
Galactic fountains may deform the \Hi disk when they fall back \citep{2017ASSL..430..323F}, of which effects may be more prominent in simulations where fountain flows extend much further out of the galaxy \citep{2019MNRAS.490.4786G,2020ApJ...895...17D} possibly due to the lack of resolution at the disk--halo interface.
Stellar winds may also change the CGM environments and thus the \Hi morphology through injecting energy and momentum, leaking ultraviolet photons, or polluting with metals.
If such processes have a short time scale compared to that of \sfr variation, these effects may cause the strong correlation between the global \sfr and \Hi irregularity or area ratios within samples of single AGN feedback mode.
In TNG50, the typical stellar winds of MW-like galaxies reach the radius around abnormal \Hi structures within \qty{\sim100}{\Myr}, taking a velocity floor of \qty{350}{\km\per\s} \citep{2018MNRAS.473.4077P}, indeed shorter than the \sfr variation timescale of \qty[input-comparators=\gtrsim]{\gtrsim300}{\Myr} found with its precursor, Illustris \citep{2015MNRAS.447.3548S}.

Another possibility is that in TNG, the stellar feedback deforms the \Hi directly and locally by generating bubbles, holes, and shells in the star-forming region \citep{2024ARA&A..62..529T}.
In TNG, this effect may be noticeable because the size of star-forming disks is on average twice as wide as in observation at a given \Mste \citep{2023ApJ...955...55W}, possibly due to the overestimation of \sfr at low gas density \citep{2003MNRAS.339..289S,2013MNRAS.436.3031V} compared with observations \citetext{\citealp{2008AJ....136.2846B,2010AJ....140.1194B,2024ApJ...973...15W}; \citealp[see also][]{2013MNRAS.436.2747K}}.
While in this picture, these oversized star-forming disks may increase the scope of stellar feedback, we do not find a correlation between the relative size of \sfr disk (normalized by \Hi disk size) and the morphological irregularity or area ratios (see \autoref{app:sec:deltaS_size_ratio}).
This lack of correlation holds in both thermal- and kinetic-mode samples, despite the latter has systematically larger morphological parameters as found in \autoref{fig:para_depend_MBH}.
Therefore, it is more likely that in TNG, the abnormal \Hi morphology is dominantly produced by the stellar feedback in an integral and global way.

As other simulations, TNG has an overly simplified star-formation model.
It uses a density threshold above which the gas is converted to stars with an efficiency determined by the Kennicutt--Schmidt relation \citep{2003MNRAS.339..289S}, without tracing the gas cooling below \qty{\sim1e4}{K} and the formation of molecular gas \citep{2009ApJ...693..216K}.
In the outer region of a real galaxy disk, the metallicity is low, \Hi is more in the warm phase than in the cool phase, and molecular gas is more difficult to form than in the inner disks \citep{2013MNRAS.436.2747K};
thus, the tension between oversimplified model and real-galaxy star formation can be even stronger in the low-\NHi region.
As mentioned at the beginning of this section, TNG also has a simplified stellar-feedback model after stars form.
In real galaxies, the feedback efficiency should depend on young stellar clustering, localized ISM pressure, structure, metallicity, dust, and gas phase \citep{2021MNRAS.504.1039R}, but these kinds of physics are currently missing in TNG\@.
These simplifications together may result in unrealistic stellar feedbacks that strongly perturb the outer \Hi disks.

Interestingly, the stellar-feedback model in TNG and Auriga is comparatively gentle among cosmological simulations \citep{2023ARA&A..61..473C}, but their galaxies are already more irregular in \Hi than observations.
It is plausible that burstier stellar feedback models are conducive to an even more irregular \Hi distribution \citep[e.g.,][]{2025MNRAS.537.3792J}, which needs to be confirmed in future work.

\subsection{Possible Mechanisms Associated with \Hi Richness}
There are at least two possible mechanisms relating the \Hi area ratio to the \Hi richness.
TNG galaxies have net in-plane radial inflow of \Hi roughly out of the optical radius \citep{2025A&A...697A..86M}, possibly transferring there the outer \Hi and decreasing area ratios.
This inflow should be stronger in a larger \Hi disk (i.e., richer in \Hi), where gas is easier to perturb and remove angular momentum.
Additionally, at the disk edge, more-extended \Hi disks in TNG should be easier to truncate by thermal and radiative ionization from the surrounding CGM \citep{2025ApJ...982..151L}.

These two mechanisms may explain the predicted anticorrelation between \MHi and \logSratOuter (lower row, \autoref{fig:spearman_r}).
However, such anticorrelation is much weaker in observation (\autoref{fig:para_depend}) and only becomes significant in less-inclined systems after correcting for projection effects \citep[see][]{2025ApJ...980...25W}, possibly being related to their much smaller \Hi disks than in simulation at the $\NHi=\qty{1e18}{\per\cm\squared}$ level (\autoref{app:sec:size_mass}).
These two mechanisms can decrease the \Hi extension, thus making simulated \Hi disks more realistic, but in TNG, they may be insufficient compared to counteracting mechanisms.
Notably, the dependence of area ratio on \MHi becomes subordinate to the one on \sfr after the AGN enters the low-accretion mode in TNG, emphasizing the complex interplay between different processes.
This change in the importance of \MHi may reflect different CGM conditions in different AGN modes, given that AGN feedback itself does not directly affect \Hi morphology (see \autoref{ssec:discuss_AGN}).

\subsection{The Auxiliary Role of AGN Feedback}
\label{ssec:discuss_AGN}
In simulations, AGN feedback has been a standard component for reproducing the bend of the \Mste--\Mhalo relation at a halo mass of \qty{\sim1e12}{\Msun} \citep{2015MNRAS.446..521S,2018MNRAS.475..648P}.
It has been highlighted that in TNG, the cumulative AGN feedback ($\propto\MBH$) increases the CGM gas entropy and cooling time \citep{2020MNRAS.499..768Z}, strongly determining the star-formation activities \citep{2022ApJ...941..205M,2022MNRAS.512.1052P}.
In this study, however, BH properties generally do not correlate with the \Hi morphology (\autoref{fig:spearman_r}), except that galaxies with abnormally irregular or extended \Hi are only found when \MBH is high or the AGN is in the low-accretion mode (Figures \ref{fig:para_depend_MBH} and~\ref{fig:deltaS_size_ratio}).

Our results favor a scenario that the AGN feedback does not influence the \Hi morphology instantly or directly.
It is more possible that it indirectly influences the \Hi via cumulatively modifying the CGM condition \citep[e.g.,][]{2022ApJ...941..205M}, not strongly dependent on specific feedback processes.
Correspondingly, when TNG galaxies transit from thermal-mode to kinetic-mode AGN feedbacks, their \Hi morphology (\autoref{fig:para_depend_MBH}) and absolute size (\autoref{fig:size_mass}) do not change abruptly, and despite having a different AGN recipe, Auriga is similar to TNG50 in \Hi morphology.
This modification of the CGM may be achieved by gradually emptying or heating the inner CGM, which in passing assists other feedback processes to reach and affect the \qty{1e18}{\per\cm\squared} \Hi gas.
On a long time scale, the metal-rich gas ejected by the kinetic-mode feedback in TNG to roughly the virial radius \citep{2024MNRAS.532.3417W} could also contribute to the \Hi reservoir, after it cools down and falls back, losing track of the instantaneous AGN feedback \citep{2019MNRAS.486.4686K}.

\section{Summary}
\label{sec:summary}
For the first time, we conduct a statistical comparison of the \Hi morphology of MW-like galaxies at a column density level close to \qty{1e18}{\per\cm\squared} between observations and cosmological simulations.
The FEASTS observation and TNG50 simulation samples are matched by stellar mass, \Hi mass, distance, and inclination.
The simulation data are mock observed and processed in a similar way as the observational data, providing a fair comparison.
We mainly use two parameters to quantify the irregularity at $\NHi\sim\qty{1e18}{\per\cm\squared}$ (\deltaS) and the spatial extension at $\NHi=\qty{1e18}{\per\cm\squared}$ (relative to \qty{1e19}{\per\cm\squared}; the area ratio \logSratOuter).

We obtain the following results:
\begin{enumerate}
  \item Compared with FEASTS, TNG50 galaxies have systematically higher levels of irregularity and extension for the \Hi gas at $\NHi=\qty{1e18}{\per\cm\squared}$ (\deltaS and \logSratOuter).
        Such an abnormality of \Hi morphology happens in ${\sim}1/3$ of our TNG50 sample.
        Their \Hi extensions at higher column density levels (spatial extension at \qty{1e19}{\per\cm\squared} relative to \qty{1e20}{\per\cm\squared}, \logSratInner) are much more similar.
        The deviation from observations mainly occurs at $\Mste\gtrsim\qty[parse-numbers=false]{10^{10.5}}{\Msun}$, $\fHi\lesssim10^{-0.5}$, or when the kinetic-mode AGN feedback is on.
        All these morphological parameters of TNG50 galaxies have a positive (negative) dependence on stellar mass \Mste (\Hi mass fraction \fHi), but those of FEASTS galaxies do not.
  \item Among the full TNG50 matched sample, the major driving property of \Hi irregularity is BH mass or stellar mass; within the kinetic-mode sample, the \sfr becomes the dominating property, and the BH mass reduces to the secondary one.
        Overall, the area ratios that reflect the \Hi radial profile extension decrease with \Hi richness; within the kinetic-mode sample, area ratios majorly increase with \sfr, and secondarily decrease with \MHi.
  \item In TNG, while the total \sfr value has a driving role on the \Hi morphological parameters, the increase in its radial extension with respect to the \Hi does not enhance the effect, suggesting its influence being global.
  \item Simulated galaxies in Auriga also have more-irregular and more-extended \Hi morphologies than observed ones in similar \MHi and \Mste ranges.
        The magnetic field in Auriga potentially has a role in making \Hi morphology more realistic, while the mass resolution does not have a clear systematic role on \Hi morphology as quantified by our parameters.
\end{enumerate}

Our comparisons highlight the difference in \Hi morphology between the FEASTS observation and TNG50 simulation, especially at a high stellar mass (\qty[parse-numbers=false]{\gtrsim10^{10.5}}{\Msun}).
Although the \Hi is usually thought to be easily perturbed in morphology, the results in this study favor a scenario that in the simulation, feedback processes only have a global effect, likely on a long time scale.
The dominant process that produces abnormal \Hi morphology in TNG50 is the stellar feedback, whose global strength has a significant correlation with \Hi morphology, while the AGN feedback is not so important as in predicting many other galaxy properties.

The \Hi radial profile shape in TNG is potentially determined by the \Hi mass (or fraction) to the first order due to synergies between mechanisms that are not fully understood at this stage.
This strong ``self-determination'' has revealed itself as the remarkably tight size--mass relation of \Hi \citep{2016MNRAS.460.2143W}, and here further as the strong anticorrelation of area ratio \logSratOuter with \Hi mass.
The in-plane radial inflow and disk transiting to the CGM may play a role, which should be more thoroughly studied in the future.

At a high column density of \qty{\sim1e20}{\per\cm\squared}, cosmological simulations are able to well reproduce \Hi properties \citep{2016MNRAS.456.1115B,2017MNRAS.466.3859M,2019MNRAS.487.1529D}, indicating the degeneracy of physical processes there.
In this study, with the \Hi depth pushed \num{\sim100} times deeper, the discrepancy in \Hi between observation and simulation becomes evident.
This study illustrates the capability of new, deep \Hi surveys to reveal the physical processes happening at the CGM--ISM interface.
Methods and parameters developed in this work could help to constrain the physics recipe of future or ongoing cosmological simulations \citep[e.g., COLd Ism and Better REsolution, or COLIBRE,][]{2025arXiv250821126S}.
While only the moment-0 map is used here, the kinematical information may provide further constraints in future studies.

\begin{figure*}
  \includegraphics{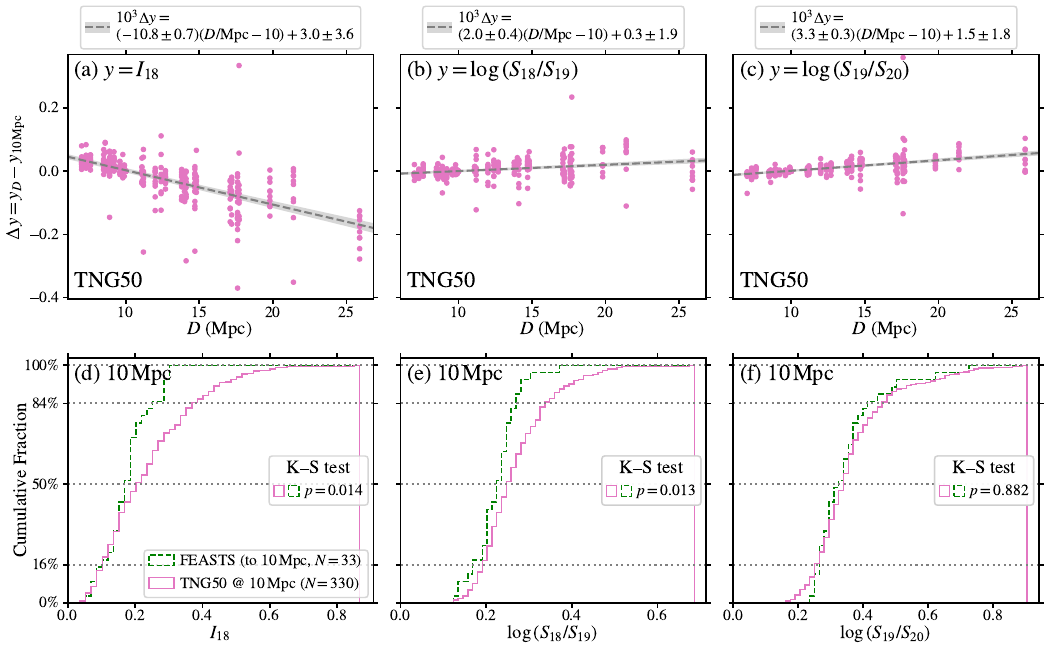}
  \caption{
    The dependence of three morphological parameters (\deltaS, \logSratOuter, and \logSratInner from left to right) and their cumulative distribution of \qty{10}{\Mpc} values.
    (a)--(c)~The difference of TNG50 distance-matched morphological parameters from their \qty{10}{\Mpc} values as a function of their distance $D$.
    The best least-squares linear fitting results are plotted as gray dashed lines.
    Uncertainties are indicated by the shading and the fitted parameters are listed above the panels.
    (d)--(f)~The same as \autoref{fig:tng_feasts_hist}, but for the \qty{10}{\Mpc} values.
    FEASTS measurements are corrected to \qty{10}{\Mpc} using the linear fitting results.
  }
  \label{fig:tng_dist_relation}
\end{figure*}

\section*{Acknowledgments}
We thank the anonymous referee for providing helpful comments.
The authors thank Dong Yang for reducing all FEASTS observational data.
We also thank Fabrizio Arrigoni Battaia, Hong Guo, Freeke van de Voort, Dandan Xu, Tianyi Yang, and Y. Sophia Dai for fruitful discussions.

J.W. acknowledges research grants and support from the Ministry of Science and Technology of the People's Republic of China (No.~2022YFA1602902), the National Natural Science Foundation of China (No.~12233001), and the China Manned Space Program (No.~CMS-CSST-2025-A08).

This work has used the data from FAST (\url{https://cstr.cn/31116.02.FAST}).
FAST is a Chinese national mega-science facility, operated by the National Astronomical Observatories, Chinese Academy of Sciences (NAOC)\@.

\facilities{FAST:500m}

\software{\softwarename{Astropy} 7.0.1 \citep{2022ApJ...935..167A}, \softwarename{MARTINI} 2.1.3 \citep{2019MNRAS.482..821O,2024JOSS....9.6860O}, \softwarename{pingouin} 0.5.5 \citep{2018JOSS....3.1026V}, \softwarename{Python} 3.12.3, \softwarename{SoFiA} 2.5.0 \citep{2021MNRAS.506.3962W}}

\appendix
\section{The Dependence of Morphological Parameters on Spatial Resolutions}
\label{app:sec:morph_dist_depend}
Spatial resolutions may influence the values of our morphological parameters.
We quantify this effect in TNG50 by measuring the parameters for both TNG50 \qty{10}{\Mpc} cubes and distance-matched cubes, and plot their differences $\inc y=y_D-y_{\qty{10}{\Mpc}}$ against the distance $D$ in \autoref{fig:tng_dist_relation}(a)--(c).
Here, we use $y$ to denote any one of three parameters, and the subscript gives the distance.
At $D=\qty{10}{\Mpc}$, $\inc y$ is \num{0} within uncertainties as expected.
The irregularity \deltaS decreases with $D$ while the two area ratios increase with $D$, consistent with the expectation that at a poorer resolution, the morphology is more biased by the shape of the beam.%
\footnote{
  One TNG50 galaxy, subhalo~494709 matched with NGC~4414, has significantly large $\inc y$ values because its \Hi tail is deblended at \qty{10}{\Mpc} but remains at $D$.
}
We linearly fit the relation between $\inc y$ and $D$ with a least-squares method, with the fitting results listed above the panels and plotted as gray dashed lines.

With these $\inc y$--$D$ correction relations from TNG50, we convert the directly measured FEASTS morphological parameters to \qty{10}{\Mpc} by subtracting the corresponding $\inc y$ value.
We plot their cumulative distributions in \autoref{fig:tng_feasts_hist}(g)--(i) along with those of the TNG50 \qty{10}{\Mpc} cubes.
Like at original distances, TNG50 galaxies still have significantly higher \deltaS and \logSratOuter than FEASTS\@.

Given the larger \Hi irregularity and spatial extension of TNG50 galaxies than FEASTS ones as shown in the main text, these fitted $\inc y$--$D$ correction relations may ``overcorrect'' the FEASTS values.
With these considerations, in the main text, we only present the comparison results between FEASTS data at original distances and the TNG50 distance-matched cubes.

\section{Measuring the Radial and Vertical Sizes of \Hi Disks}
\label{app:sec:size_mass}
\Hi disks have been found to follow tight size--mass relations with both \Ri and \Rooi, the radii at \Hi surface densities of \qtylist{1;0.01}{\Msun\per\pc\squared} \emph{after} correcting for projection effects \citep{1997A&A...324..877B,2016MNRAS.460.2143W,2025ApJ...980...25W}.
We measure \Ri and \Rooi using the same method as \citet{2025ApJ...980...25W} with \qty{10}{\Mpc} cubes, accounting for both the projection and beam-smoothing effects.

As seen in \autoref{fig:size_mass}(a) and~(b), for \Ri, TNG50 generally follows the observational trend.
Auriga systematically overpredicts \Ri at a given \MHi, but the extent is small compared to the scatter of the relation.
For \Rooi, both TNG50 and Auriga have a significant fraction of population above the observational trend, consistent with our results that area ratios are higher in simulations.
Meanwhile, for TNG, the overextended galaxies can be found in both thermal- and kinetic-mode sample, hinting a smooth transition in CGM or \Hi morphological parameters between two AGN-feedback modes.

In the direction perpendicular to the disk, \Hi structures and extensions are possible indicators of the gas circulation in feedback-driven fountains.
For seven edge-on FEASTS galaxies, \citet{2025ApJ...984...15Y} measured their individual average \zIB's, which show a strong positive relation with \MHi (black line in \autoref{fig:size_mass}(c)).
Observed \zIB's are significantly smaller than the statistical values in Illustris or TNG100 \citep{2016MNRAS.462.3751K,2019MNRAS.486.4686K} measured from the average vertical neutral-hydrogen profile within a sample spanning a wide \fHi range.

To provide a more detailed comparison, we measure the \zIB values of our TNG50 matched sample with a method similar to that in \citet{2025ApJ...984...15Y}, but we effectively measure the maximum vertical extension of a galaxy considering the irregularity of TNG50 \Hi disks, instead of its individually averaged one.
For FEASTS galaxies, the maximum and individually averaged \zIB values show no difference due to their regular morphology.
We generate their edge-on moment-0 maps at \qty{10}{\Mpc} without smoothing by a beam, clip the pixels above \qty{1e19} {\per\cm\squared} to \qty{1e19}{\per\cm\squared} to reduce the scatter light, then convolve the maps with the FAST beam.
We calculate the average \NHi profiles perpendicular to the disk in each consecutive \qty{10}{\kpc}-wide bin (${\sim}\fwhm$) within the central $2\Ri$ along the major axis, and get \zIB's by interpolation.
The maximum values of \zIB in each side are taken as the two values of \zIB for this galaxy, plotted as crosses in \autoref{fig:size_mass}(c).

\begin{figure}
  \centering
  \includegraphics{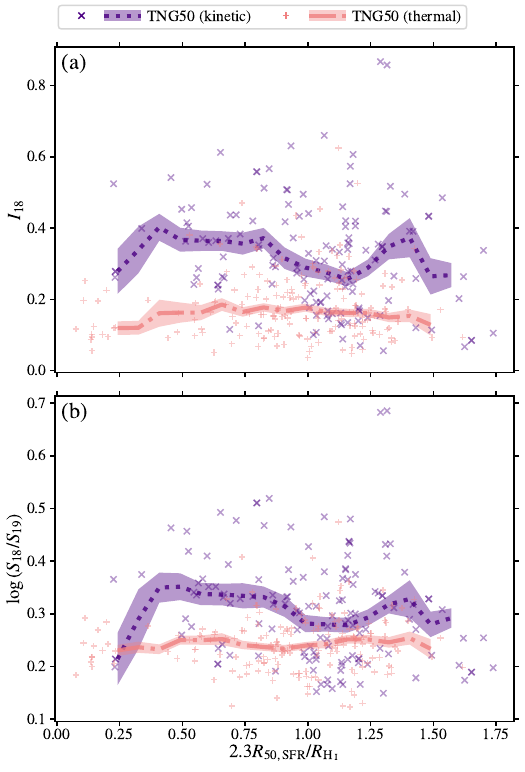}
  \caption{
    Morphological parameters of the TNG50 matched sample compared with their relative \sfr-disk size normalized by \Hi-disk size.
    We take \num{2.3} times the half-\sfr radius as the radius enclosing \qty{90}{\percent} of \sfr, assuming an exponential \sfr disk.
    The symbols are the same as those in \autoref{fig:para_depend_MBH}.
  }
  \label{fig:deltaS_size_ratio}
\end{figure}

Different from the observations, the \zIB in TNG50 shows no clear correlation with \MHi.
Meanwhile, our TNG50 \zIB are somewhat smaller than the values from \citep[\qty{\sim100}{\kpc},][]{2019MNRAS.486.4686K}, possibly due to the differences in \fHi and \Mste range and \Hi postprocessing.

\section{The Role of \sfr-disk Size}
\label{app:sec:deltaS_size_ratio}
In \autoref{fig:deltaS_size_ratio}, we compare the morphological parameters of the TNG50 matched sample with their ratio between the radii of \sfr and \Hi disk.
Assuming that the \sfr disk has an exponential profile, we take \num{2.3} times the half-\sfr radius \citep[$R_{50,\sfr}$, from][]{2018MNRAS.474.3976G}%
\footnote{
  \url{https://www.tng-project.org/data/docs/specifications/\#sec5e}
}
as the radius enclosing \qty{90}{\percent} of \sfr, and normalize it by \Ri.
No correlation is found between morphological parameters and the normalized \sfr-disk size.

\bibliography{ms}

\end{document}